\documentclass[twocolumn]{aastex62}
\usepackage{amsmath}
\usepackage{graphicx}
\usepackage[normalem]{ulem}
\usepackage{natbib}

\graphicspath{{./}{figs/}}

\newcommand\redout{\bgroup\markoverwith
{\textcolor{red}{\rule[.5ex]{2pt}{0.4pt}}}\ULon}
\usepackage{xcolor}
\usepackage{threeparttable}

\newcommand{\zfrb}{0.3216}

\newcommand{\mzzfrb}{z_{\rm host}}
\newcommand{\zzfrb}{$\mzzfrb$}
\newcommand{\mzphot}{z_{\rm phot}}
\newcommand{\zphot}{$\mzphot$}

\newcommand{\mdmunits}{{\rm pc \, cm^{-3}}} 
\newcommand{\dmunits}{$\mdmunits$}
\newcommand{\mdmcosmic}{{\rm DM}_{\rm cosmic}}
\newcommand{\dmcosmic}{$\mdmcosmic$}
\newcommand{\mdmacosmic}{\langle {\rm DM}_{\rm cosmic} \rangle}
\newcommand{\mdmahalos}{\langle {\rm DM}_{\rm halos} \rangle}
\newcommand{\mdmaigm}{\langle {\rm DM}_{\rm IGM} \rangle}
\newcommand{\dmacosmic}{$\mdmacosmic$}
\newcommand{\dmahalos}{$\mdmahalos$}
\newcommand{\dmaigm}{$\mdmaigm$}

\newcommand{\mdmfrb}{{\rm DM}_{\rm FRB}}
\newcommand{\dmfrb}{$\mdmfrb$}
\newcommand{\mdmigm}{{\rm DM}_{\rm IGM}}
\newcommand{\dmigm}{$\mdmigm$}
\newcommand{\mdmhalos}{{\rm DM}_{\rm halos}}
\newcommand{\dmhalos}{$\mdmhalos$}
\newcommand{\mdmhost}{{\rm DM}_{\rm host}}
\newcommand{\dmhost}{$\mdmhost$}
\newcommand{\mdmmw}{{\rm DM}_{\rm MW}}
\newcommand{\dmmw}{$\mdmmw$}

\newcommand{\mdmhalosphot}{\rm DM_{phot,halos}}
\newcommand{\dmhalosphot}{$\mdmhalosphot$}
\newcommand{\mdmphotgal}{\rm DM_{phot,galaxy}}
\newcommand{\dmphotgal}{$\mdmphotgal$}

\newcommand{\mmstar}{M_*}
\newcommand{\mstar}{$\mmstar$}

\newcommand{\mmsun}{{\rm M}_\odot}

\newcommand{\mmhalo}{M_{\rm halo}}
\newcommand{\mhalo}{$\mmhalo$}

\submitjournal{ApJ}

\shorttitle{\dmhalos~in the FRB 180924 sightline}
\shortauthors{Simha et al.}

\begin{document}

\title{Estimating the contribution of foreground halos to the FRB 180924 dispersion measure}

\correspondingauthor{Sunil Simha}
\email{shassans@ucsc.edu}

\author[0000-0003-3801-1496]{Sunil Simha}
\affil{University of California - Santa Cruz
1156 High St.
Santa Cruz, CA, USA 95064}

\author[0000-0002-1883-4252]{Nicolas Tejos}
\affil{Instituto de F\'isica, Pontificia Universidad Cat\'olica de Valpara\'iso, Casilla 4059, Valpara\'iso, Chile}

\author[0000-0002-7738-6875]{J. Xavier Prochaska}
\affil{University of California - Santa Cruz
1156 High St.
Santa Cruz, CA, USA 95064}
\affil{Kavli IPMU (WPI), UTIAS, The University of Tokyo, Kashiwa, Chiba 277-8583, Japan}

\author[0000-0001-9299-5719]{Khee-Gan Lee}
\affil{Kavli IPMU (WPI), UTIAS, The University of Tokyo, Kashiwa, Chiba 277-8583, Japan}

\author[0000-0003-4501-8100]{Stuart D. Ryder}
\affil{Department of Physics \& Astronomy, Macquarie University, NSW 2109, Australia}
\affil{Macquarie University Research Centre for Astronomy, Astrophysics \& Astrophotonics, Sydney, NSW 2109, Australia}

\author[0000-0001-5804-1428]{Sebastiano Cantalupo}
\affil{Department of Physics, ETH Zurich, Wolfgang-Pauli-Strasse 27, 8093, Zurich, Switzerland}
\affil{Department of Physics, University of Milan Bicocca, Piazza della Scienza 3, 20126 Milano, Italy}

\author[0000-0003-2149-0363]{Keith W. Bannister}
\affil{CSIRO, Space and Astronomy, PO Box 76, Epping NSW 1710 Australia}

\author[0000-0003-3460-506X]{Shivani Bhandari}
\affil{CSIRO, Space and Astronomy, PO Box 76, Epping NSW 1710 Australia}

\author[0000-0002-7285-6348]{Ryan M. Shannon}
\affil{Centre for Astrophysics and Supercomputing, Swinburne University of Technology, Hawthorn, VIC 3122, Australia}




\begin{abstract}
Fast Radio Burst (FRB) dispersion measures (DMs) record the presence of ionized baryons that are otherwise invisible to other techniques enabling resolution of the matter distribution in the cosmic web. In this work, we aim to estimate the contribution to FRB 180924 DM from foreground  galactic halos. Localized by ASKAP to a massive galaxy, this sightline is notable for an estimated cosmic web contribution to the DM (\dmcosmic = 220 \dmunits), which is less than the average value at the host redshift (z = \zfrb) estimated from the Macquart relation (280 \dmunits). In the favored models of the cosmic web, this suggests few intersections with foreground halos at small impact parameters ($\lesssim 100$ kpc). To test this hypothesis, we carried out spectroscopic observations of the field galaxies within $\sim$1\arcmin\ of the sightline with VLT/MUSE and Keck/LRIS. Furthermore, we developed a probabilistic methodology that leverages photometric redshifts derived from wide-field DES and WISE imaging. We conclude that there is no galactic halo that closely intersects the sightline and also that the net DM contribution from halos, \dmhalos\ $< 45~\mdmunits$ (95 \% c.l.). This value is lower than the 
\dmhalos\ estimated from an ``average'' sightline (121 \dmunits) using the Planck $\Lambda$CDM model and the Aemulus halo mass function and reasonably explains its low \dmcosmic\ value.  We conclude that FRB 180924 represents the predicted majority of sightlines in the universe with no proximate foreground galactic halos.  Our framework lays the foundation for a comprehensive analysis of FRB fields in the near future.
\end{abstract}


\section{Introduction}
\label{sec:intro}
Fast Radio Bursts (FRBs) are millisecond-duration, energetic
($\sim10^{44}$ erg) radio transient events. In recent years, numerous FRBs have been localized and most FRBs are confirmed to be extragalactic \citep[e.g.][]{lorimer2007, Bannister+19, tendulkar+2017, law+2020}. Although their generation mechanism is yet unknown, FRBs represent a new tool in the repertoire of an observational cosmologist to probe matter and cosmological structure in the universe. Astronomers have used quasar absoprtion lines to study neutral gas in the circumgalactic medium \citep[e.g.][]{bs69, ct08, prochaska+11, tumlinson+13, cos_halos, wilde+21} for the last several decades. Now, with their unique transient signal, FRBs enable us to capture information about all ionized matter along their lines-of-sight, thus unlocking an opportunity to study previously invisible gas in the universe. One of the measurable properties of FRBs is their dispersion measure (DM), which is the cosmological-scale-factor-weighted line of sight integral of electron density. Even with a handful of localized FRBs, \citet{macquart2020} were able to show that the observed FRB DMs are consistent with the expected matter distribution in a $\Lambda$CDM universe, thus conclusively resolving the Missing Baryon Problem. While their work ``found'' the Missing Baryons, the next phase of research is to precisely locate them within the cosmic web. Specifically, we aim to develop the framework to utilize these data and reconstruct the distribution of matter along the sightlines.

Being an integral, one can split the FRB DM into disjoint summative parts corresponding to each ``electron reservoir'' along the line of sight, namely the host galaxy and its halo; intervening foreground halos and cosmic web filaments; and the Milky Way including its gaseous halo, i.e.
 \begin{equation}
     \mdmfrb = \mdmhost+\mdmcosmic+\mdmmw
 \end{equation}
 A full characterization of \dmfrb~requires detailed information on the host \citep[e.g.][]{chittidi+20} and the intervening cosmic web structures. 
\citet{Simha+20} performed such an analysis on the sightline of FRB~190608 owing to the favorable location of the FRB in the SDSS spectroscopic footprint. This provided detailed information on the redshifts of foreground galaxies, allowing for a nearly complete characterization of their DM contributions (although with significant uncertainty) and also the contribution from the diffuse intergalactic medium (IGM). In general, if a FRB host is located at low redshifts ($z<0.05$), one could use just the 2MASS Redshift Catalog \citep[][]{2mass} to perform the same analysis. However, the vast majority of localized FRBs to date fall outside the extant SDSS coverage, and therefore one would require extensive new spectroscopic observations. In this work, we explore the application  of photometric redshifts combined with sparse spectroscopy to estimate the DM contributions of foreground halos, \dmhalos, for one such sightline: FRB 180924.

FRB 180924 was the first apparently non-repeating FRB to be discovered and localized by the Australian Square-Kilometer Array Pathfinder (ASKAP) in September 2018 \citep{Bannister+19} with a measured \dmfrb~= 362.16 \dmunits. Its massive, moderately star-forming host galaxy ($z$ = \zfrb) is located in the footprint of the first data release of the Dark Energy Survey \citep[DES DR1][]{des}. While its \dmhost~is uncertain, if one assumes it to be 66 \dmunits~ in the host rest frame \citep{Bannister+19}; uses the NE2001 model \citep{ne2001} for the Milky Way disk (41 \dmunits); and 50 \dmunits\ for the Milky Way halo gas, the remainder of the DM
is attributed to the cosmic web:
$\rm DM _{\rm cosmic, FRB} \approx 220~$\dmunits. 
This is lower than the mean expected \footnote{Expectation value obtained by assuming a flat, $\Lambda$CDM cosmology with Planck 2015 parameters and that the diffuse gas is fully ionized. The fraction of baryons in the universe constituting diffuse gas is obtained by subtracting the fractions for the dense components: stars \citep{Fukugita04} and the ISM \citep{MadauDickinson14}.} \dmcosmic~value at the host redshift (\dmacosmic = 280 \dmunits), 
suggesting either a less than average foreground matter density, or our adopted values for \dmhost\ or \dmmw\ are too large. Of course, the average value is not representative of all sightlines and there is naturally some scatter \citep[][]{macquart2020}. For a given redshift, the distribution of \dmcosmic~is skewed towards lower than average DM. This is because most of the sightlines in the universe rarely intersect any galactic halo at low impact parameter ($\lesssim50$ kpc). In this work, we test whether the lower than average \dmcosmic~is consistent with this paradigm, i.e. if there are indeed no foreground galactic halos in close proximity to it. We also outline a framework to estimate, based on photometry alone, the halo contribution to \dmfrb: \dmhalos.  

This paper is organized as follows: we describe the data collected in section \ref{sec:data}, our methods in estimating \dmhalos in section \ref{sec:methods}, our results in section \ref{sec:results} and make concluding remarks in section \ref{sec:discuss}. We assume a $\Lambda$CDM cosmology with the cosmological parameters derived from the 2015 Planck dataset \citep{Planck15} for all our calculations.

\section{Data}
\label{sec:data}
\subsection{Photometry}
We obtained photometric data in the $griz$Y bands for all sources within 
$15'$ of FRB~180924 from DES DR1 \citep[][95\% complete to $r=23.35$]{des}. This was supplemented with photometry from the Wide-field Infrared Survey Explorer \citep[WISE;][]{wise} database where available. 
At $z=0.03$, $15'$ corresponds to 560 kpc in projected physical distance. This is approximately the virial radius of a $10^{13}~\mmsun$ dark matter halo with a modified NFW profile \citep{xyz19}. Thus we hoped to capture all galaxy halos that are less massive than this limit at $z > 0.03$. We did not find any galaxy cluster or group catalog that covers this FRB sightline either and so our analysis is blind to halos of that mass scale.

To remove stars from the photometric catalog, we used the morphology-based classifier flag \texttt{class\_star\_r} from the DES DR1 database. Extended objects like galaxies have flag values closer to zero while point sources tend to lie closer to unity. We excluded objects whose $r$-band magnitudes were less than 17 and the flag value was above 0.9. To further exclude stars, we cross matched the remaining DES objects with stars having measured parallaxes (\texttt{parallax\_over\_error}$>1$) in the main GAIA DR3 catalog \citep{gaia, gaia_dr3}. The DES (and WISE) magnitudes obtained are the elliptical aperture magnitude based on the Kron radius (i.e. the \texttt{auto\_mag} columns in the main DES DR1 catalog; see Table \ref{tab:photom}).
\begin{table*}
\centering
\caption{\footnotesize Catalog of photometry from DES and WISE for galaxies in the FRB 180924 field. \tablenotemark{$\dagger$}\label{tab:photom}}
\begin{tabular}{|c|c|c|c|c|c|c|c|c|c|c|c|c|}
\hline
DES ID\tablenotemark{a} & RA & Dec & g & r & i & z & Y & W1 & W2 & W3 & W4 & \zphot \\
 & $\deg$ & $\deg$ & mag & mag & mag & mag & mag & mag & mag & mag & mag &  \\
\hline
209914488\tablenotemark{b} & 326.10521 & -40.90023 & 21.62 & 20.54 & 20.14 & 19.85 & 19.81 & 16.85 & 16.06 & 11.69 & 8.50 & 0.321 \\
209914542 & 326.10163 & -40.89981 & 25.05 & 24.54 & 23.92 & 23.46 & 24.77 & \nodata & \nodata & \nodata & \nodata & 0.998 \\
209914588 & 326.11102 & -40.90060 & 23.19 & 22.63 & 22.46 & 22.14 & 23.08 & \nodata & \nodata & \nodata & \nodata & 0.480 \\
209914804 & 326.10812 & -40.90395 & 26.72 & 24.57 & 23.60 & 22.91 & 22.08 & 16.68 & 15.98 & 12.20 & 8.49 & 1.391 \\
209914529 & 326.11133 & -40.89956 & 24.38 & 24.29 & 23.57 & 23.61 & 22.95 & \nodata & \nodata & \nodata & \nodata & 0.963 \\
\hline
\end{tabular}
\hspace{-1.7in}
\begin{flushleft}
\tablenotetext{\dagger}{\footnotesize This table is published in its entirety in the machine-readable format. Five galaxies with the lowest angular separation to the FRB sightline are shown here. The last column is the best photometric redshift derived using EAZY (see section \ref{sec:prelim_analysis}).}
\tablenotetext{\rm a}{\footnotesize \texttt{COADD\_OBJECT\_ID} from DES DR1 Main.}
\tablenotetext{\rm b}{\footnotesize Host galaxy.}
\end{flushleft}
\end{table*}

\subsection{Spectroscopy}
\label{sec:spectra}
\begin{figure*}
\includegraphics[width=\textwidth,trim={0.25in 0.2in 0.4in 0.2in},clip]{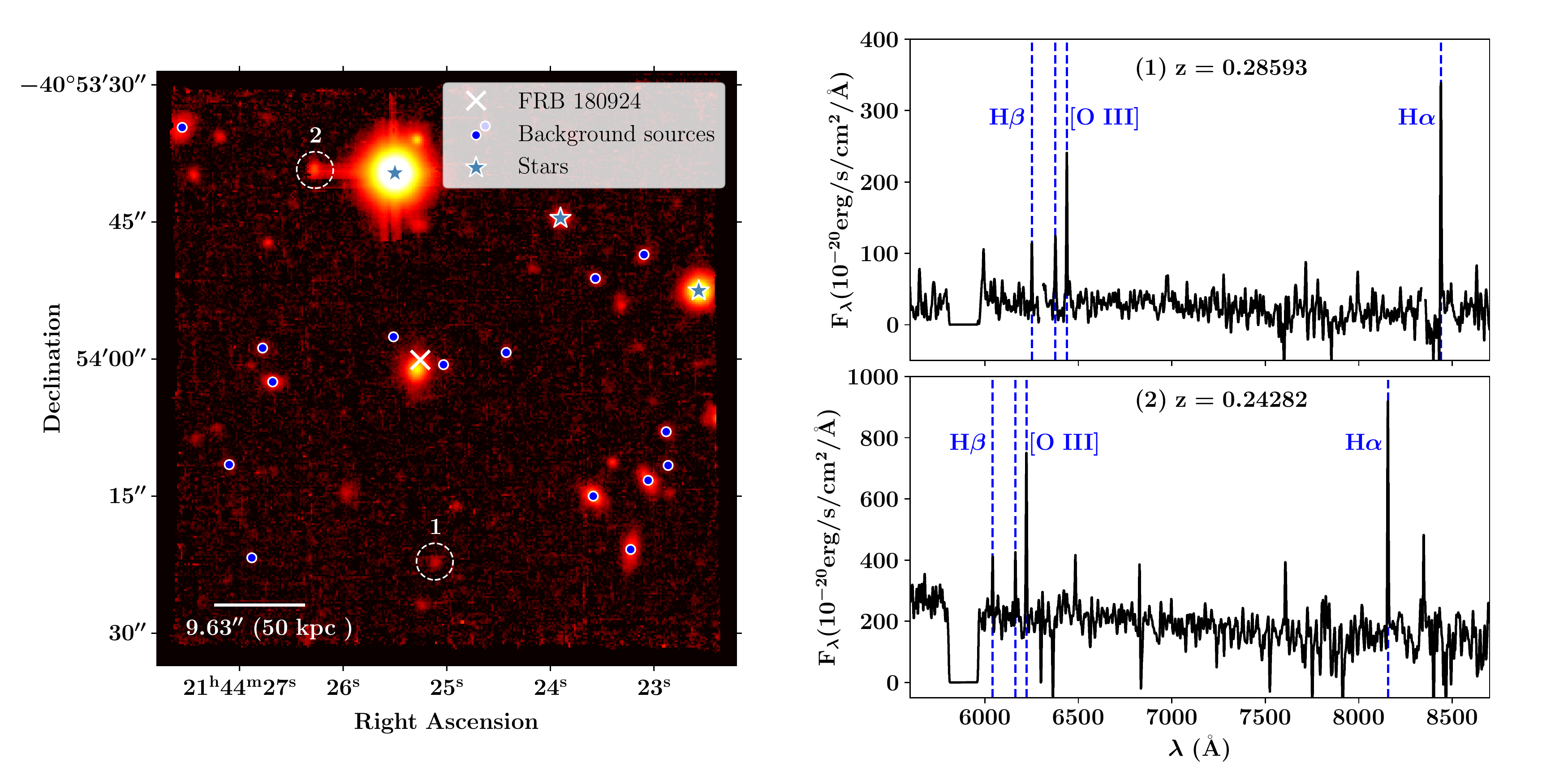}
\caption{\footnotesize \textit{Left:} MUSE white light image of the $1\arcmin \times 1\arcmin$ field around the host galaxy of FRB 180924. The circled objects are the two foreground galaxies and the blue dots are background sources identified using MARZ. The blue stars mark the stars in the field. The redshifts of the unmarked sources in the image could not be identified due to their spectra either being noisy or not having a clear correlation with any of the default MARZ template spectra. The white bar at the bottom represents 50 kpc at \zzfrb. \textit{Right:} The spectra of sources 1 and 2. Both galaxies show clear H$\alpha$, H$\beta$ and [O III] doublet emission lines which help pinning their redshifts to the values noted above the spectra.}
\label{fig:muse_imgspec}
\end{figure*}

For the galaxies at $0.001<z<0.05$ along the sightline, we turned to the 2MASS survey database \citep{2mass} for spectroscopic data. This catalog contains galaxy spectra of 83\% of the southern sky and is complete to $J < 13.75$ with median redshift $z=0.053$. 
We determined that the galaxy in this catalog with the smallest perpendicular distance to the FRB sightline was 1.04 Mpc (15.2$\arcmin$ angular distance) away, which is far beyond the typical virial radii of galaxy halos. We also found no galaxy within 500 kpc in NearGalCat, the updated nearby galaxy catalog of 869 galaxies within 11 Mpc of the Milky Way which is estimated to be $\sim40\%$ complete \citep{neargalcat}. Thus, we conclude the $z<0.03$ intervening galaxy halo contribution to \dmfrb\ is negligible if not null.

To survey galaxies close to the FRB sightline we used the MUSE integral field unit \citep{muse} on the Very Large Telescope (VLT). A set of $4\times 628$\,s exposures were obtained on UT 2018 November 5 from program 2102.A-5005 (PI Macquart); another set of $4\times 600$\,s exposures were obtained on UT 2019 December 6 from program 0104.A-0411 (PI Tejos). These observations were carried out in the Wide-Field Adaptive Optics (WFM-AO) mode, corresponding to a effective FoV of $1\arcmin \times 1\arcmin$ with a pixelscale of $0.2\arcsec$, and covering a wavelength range of $R \approx 4800-9300$\,\AA\ at a resolving power of $\sim 2000-4000$, respectively.
After preliminary reduction using the EsoReflex pipeline \citep{esoreflex}, the frames were flat-field corrected, sky-subtracted and co-added using the CubExtractor package \citep[see ][ for a description]{cubex}. Sources in the datacube were identified from the white light image, i.e., the cube collapsed along the spectral dimension (see Figure \ref{fig:muse_imgspec}), using the Source Extractor of Python (SEP) package \citep{SExtractor,Barbary2016}. We set  minimum threshold of 3 standard deviations above the sky background level and with a minimum area of 10 pixels. Their spectra were extracted from the spaxels within the elliptical apertures whose linear dimensions were twice as large as those returned by SEP. The extraction weighted the flux from the spaxels encircled by the aperture equally. Where the aperture intersected a spatial pixel, the flux from that pixel was scaled down by the fraction of the pixel area within the aperture. Redshifts were identified for each spectrum from the emission features using the Manual and Automatic Redshifting software \citep[MARZ;][]{marz}. Out of the 72 non-stellar sources identified from the white light image, 19 had their redshifts confidently assigned. These objects are listed in Table \ref{tab:muse_gals}.  
The remaining spectra did not have identifiable spectral features (e.g. emission lines). Further relaxing the source detection criteria for SEP increased the number of ``sources" but did not increase the number of identified redshifts. 
From the secure redshifts, we identify two foreground sources from the datacube ($z=0.24282$ and $0.28593$; see Figure \ref{fig:muse_imgspec}). Only the closer galaxy is detected in the DES $grizY$ imaging catalog. 

\begin{table*}
    \centering
    \caption{\footnotesize MUSE and LRIS sources with unambiguous redshifts.\label{tab:muse_gals}}
    \begin{tabular}{|c|c|c|c|c|c|c|}
    \hline
    RA & Dec & Redshift & DES ID\tablenotemark{a} & r & Separation\tablenotemark{b} & Transverse distance\tablenotemark{c}\\
    $\deg$ & $\deg$ &  &  & mag & arcmin & kpc \\
    \hline
    326.10947 & -40.89425 & 0.24282 & 209914195 & 22.71 & 0.40 & 94 \\
    326.10465 & -40.90616 & 0.28593 & \nodata & \nodata & 0.37 & 98 \\
    326.10538\tablenotemark{d} & -40.90030 & 0.32157 & 209914488 & 20.54 & 0.01 & 3 \\
    326.10430 & -40.90017 & 0.38406 & \nodata & \nodata & 0.04 & 13 \\
    326.11117 & -40.90069 & 0.38407 & 209914588 & 22.63 & 0.26 & 85 \\
    326.09677 & -40.90579 & 0.46956 & 209914896 & 22.09 & 0.52 & 189 \\
    326.10631 & -40.89932 & 0.50086 & \nodata & \nodata & 0.06 & 24 \\
    326.09828 & -40.90417 & 0.53556 & 209914807 & 21.87 & 0.40 & 158 \\
    326.09607 & -40.90369 & 0.54431 & 209914777 & 22.41 & 0.47 & 186 \\
    326.09527 & -40.90323 & 0.54464 & \nodata & \nodata & 0.49 & 193 \\
    326.11480 & -40.89296 & 0.61709 & 209914131 & 22.10 & 0.60 & 252 \\
    326.10178 & -40.89980 & 0.75084 & 209914542 & 24.54 & 0.16 & 74 \\
    326.09534 & -40.90221 & 0.75097 & 209914676 & 23.54 & 0.47 & 213 \\
    326.10262 & -40.89292 & 0.86432 & \nodata & \nodata & 0.44 & 210 \\
    326.09819 & -40.89755 & 0.87372 & 209914406 & 23.86 & 0.36 & 171 \\
    326.11157 & -40.89967 & 1.03010 & 209914529 & 24.29 & 0.28 & 138 \\
    326.09623 & -40.89683 & 1.03477 & 209914359 & 24.06 & 0.46 & 228 \\
    326.11291 & -40.90321 & 1.43899 & \nodata & \nodata & 0.40 & 206 \\
    326.11201 & -40.90604 & 2.95747 & \nodata & \nodata & 0.47 & 225 \\
    326.09791\tablenotemark{e} &	-40.93261 & 0.07221 & 1.98 & 209916475 & 18.69 & 169 \\
    \hline 
    \end{tabular}
    \hspace{-1.7in}
    \begin{flushleft}
    \tablenotetext{\rm a}{\footnotesize \texttt{COADD\_OBJECT\_ID} from DES DR1 Main. Ellipses imply objects are not present in the DES database.}
    \tablenotetext{\rm b}{\footnotesize Angular separation from the FRB.}
    \tablenotetext{\rm c}{\footnotesize Angular separation converted to projected distance from the FRB sightline.}
    \tablenotetext{\rm d}{\footnotesize Host galaxy.}
    \tablenotetext{\rm e}{\footnotesize LRIS spectrum.}
    \end{flushleft}
\end{table*}

We also obtained spectra of 5 galaxies using the Low Resolution Imaging Spectrograph (LRIS) installed on the Keck telescope on November 8 2020 in the longslit spectroscopy mode. These galaxies were targeted as our analysis indicated they could contribute to \dmfrb (see Section \ref{sec:halo_dm}). We used the ``d560" dichroic, 600/4000 grism on the blue side and 600/7500 grating for the red side with $2 \times 2$ binning on both detectors. Three of these spectra were exposed for $\sim700$ s while the other two were exposed for $\sim300$\,s. We could not expose longer on account of bad weather. Of the 5 galaxies, one was confidently assigned a redshift using MARZ and it was determined to be a foreground source ($z=0.07221$; 169 kpc away). We did not detect identifiable emission lines in the remaining 4 low S/N spectra.

\section{Methods}
\label{sec:methods}
\subsection{Photometric analysis}
\label{sec:prelim_analysis}
We aim at estimating the DM contribution of galaxies that only have photometric redshifts, for which we require several intermediate derived quantities to then compute \dmhalosphot, the DM contribution of galaxies without spectroscopic redshifts in our sample. Namely, we require photometric redshifts, \zphot, and halo mass estimates, \mhalo, for every galaxy.

We first estimated the posterior distribution of \zphot~for each DES galaxy using the EAZY software package \citep{eazy}. Redshifts were only computed for those galaxies which were detected in at least four of the nine filters considered (five from DES and four from WISE) and were estimated in a Bayesian framework using template spectral energy distribution (SED) fitting. We used linear combinations of the templates available in the \texttt{eazy\_v1.3} set and applied magnitude priors on the $r$-band photometry when available \citep[see details in ][]{eazy}. When fitting, the redshift was allowed to freely vary between 0.01 and 7 but the priors heavily penalized redshifts higher than 2.

The estimation of halo masses is less direct. Briefly, starting with an estimate for the galaxy's redshift based on the photometry, we fitted the available fluxes with an SED using the CIGALE software package \citep{cigale, Boquien+2018}. We assumed, for simplicity, a delayed-exponential star-formation history with no burst population, a synthetic stellar population prescribed by \citet{bc03}, the \citet{Chabrier03} initial mass function (IMF), dust attenuation models from \citet{calzetti01}, and dust emission templates from \citet{dale14}, where the AGN fraction was capped at 20\%. This provided an estimate of the stellar mass, \mstar, of the galaxy at a given $z$. We then translate \mstar~to galactic halo mass, \mhalo, using the mean Stellar to Halo Mass Ratio (SHMR) described by \citet{moster+13} at that $z$. For sources with spectroscopic redshifts, the galaxy redshift is fixed in the CIGALE input. We elaborate on the use of \zphot~posteriors for the remaining sources in the next subsection.

\begin{figure*}[!ht]
    \centering
    \includegraphics[width=\textwidth]{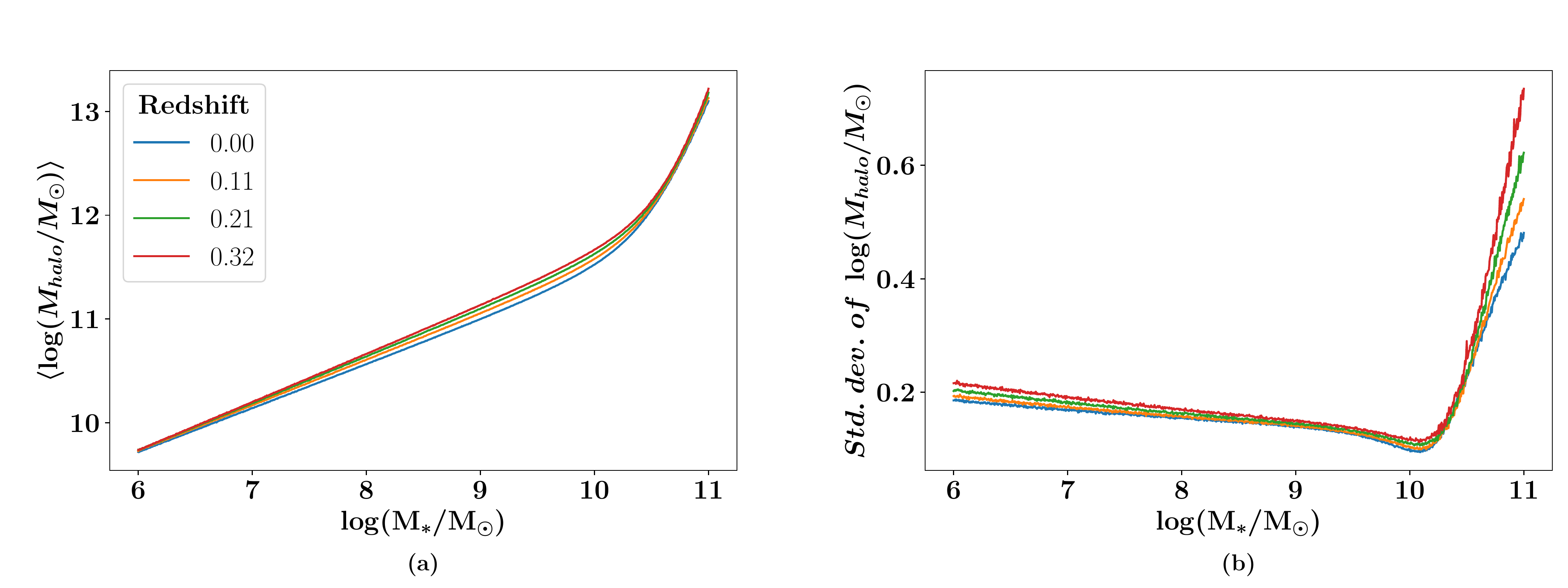}
    \caption{\footnotesize \mhalo~stellar to halo mass ratio (SHMR) mean and standard deviation obtained from sampling the fit parameter space from \citet{moster+13}. The SHMR relation (their eq. 2) contains 8 fit parameters and in this work, we have assumed they are independent and normally distributed. The mean and standard error of these fit parameters were obtained from their Table 1. Using these curves, interpolation functions are constructed to translate (\mstar,z) pairs to \mhalo~ distributions.}
    \label{fig:mhalo_input}
\end{figure*}

The uncertainties in the \mstar~estimation and the SHMR relation propagate into the \dmhalos~estimate. For each galaxy, we assumed that the $\log\mmstar$ distribution at a given redshift was Gaussian with the mean and standard deviations obtained from CIGALE. Accounting for the error in the SHMR is more involved as it depended on both \mstar~and galaxy redshift. The SHMR is described in Equation 2 of \citet{moster+13} with 8 parameters. We took the best fit parameters and uncertainties from their Table 1 as the mean and standard deviations of the independent normal distributions that these parameters were sampled from. For simplicity, we ignored any co-variance in these fit parameters (future work will account for this). We then produced a uniform 2D grid of redshift (between 0.03 and 0.35 spaced by 0.01) and $\log\mmstar/\mmsun$ (between 6 and 11 spaced by 0.005). At each grid point, we sampled the parameter distributions and produced a lookup table of the mean and standard deviations of halo masses that can be realized. Then, to quicken computation, we constructed interpolation functions that mapped the 2D grid to the mean and standard deviation of $\log\mmhalo/\mmsun$. Figure \ref{fig:mhalo_input} shows the mean and standard deviations for some representative redshift and stellar mass values. The halo mass distributions were assumed to be Gaussian with the moments given by these interpolation functions.

\subsection{Halo contribution to DM}
\label{sec:halo_dm}

To estimate \dmhalos, we performed an analysis similar to the one outlined by \citet{Simha+20} for the  FRB 190608 sightline. Briefly, they identified foreground galaxies based on spectroscopic redshifts and estimated halo masses from the available photometry. Then they estimated the line of sight electron number density integral for each intevening halo assuming a model for the baryonic distribution and summed the contributions to yield \dmhalos. We emphasize that the redshift serves as a key input to each step of the analysis.
In the case of FRB 180924, we modified the procedure to leverage galaxies with \zphot~as follows (see Figure \ref{fig:schematic} for a visual flowchart):

\begin{figure*}[ht!]
    \centering
    \includegraphics[width=\textwidth]{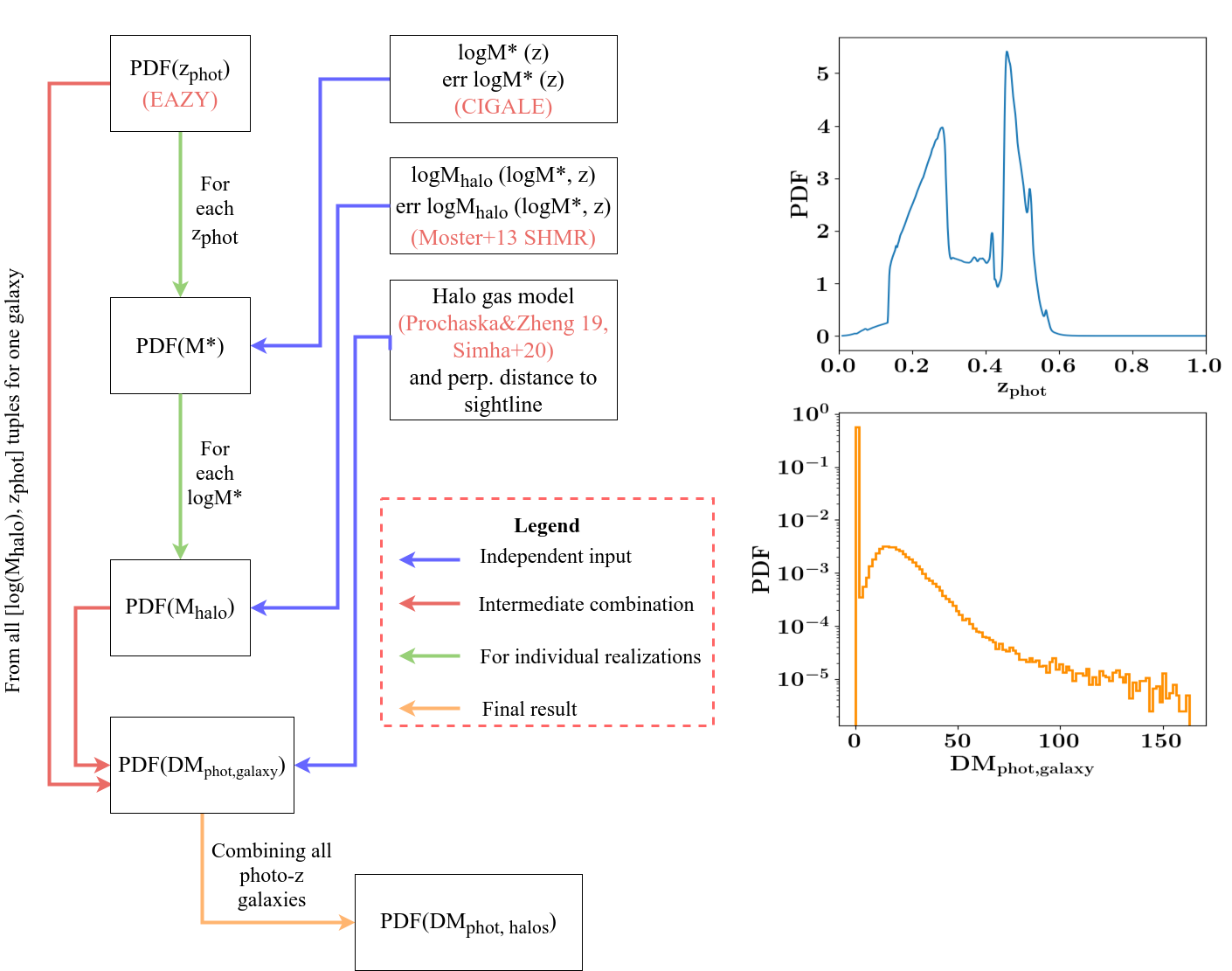}
    \caption{\footnotesize A schematic flowchart of our procedure to estimate \dmhalosphot. The boxes in the centre with blue arrows emanating from them represent independent inputs into the calculation. These include the stellar mass estimates, the SHMR and the halo gas model. The \zphot~PDF is also an independent input and an example for one galaxy is shown on the plot on the top right. The sources of these estimates are mentioned in red lettering. The PDF of \dmhalosphot\ is obtained in stages. First, the PDF of stellar masses at each redshift (sampled from the EAZY \zphot~posterior) is obtained. Then each stellar mass and redshift tuple is translated to halo mass distributions using the \citet{moster+13} SHMR relation. Compiling all the halo mass and redshift tuples and calculating their DM contribution \citep[using the method outlined in ][]{xyz19,Simha+20}, yields a PDF of DM values for each individual galaxy in our sample. An example of this is shown in the bottom right plot for the same galaxy as the \zphot~PDF plot. The final PDF of \dmhalosphot\ is estimated by sampling the galaxy DM PDFs and obtaining the distribution of the sum of these samples.}
    \label{fig:schematic}
\end{figure*}

For a given galaxy:
\begin{enumerate}
    \item We estimated the posterior distributions for \zphot\ and sampled them to produce 1000 realizations.
    \item Separately, we allowed the galaxy redshift to vary from 0.03 to 0.35~in a linear grid (spacing 0.01) and estimated the mean and standard deviation of the stellar mass at each grid point using CIGALE.
    \item Then, at each redshift realization from step 1, we sampled the $\log\mmstar$ distribution (100 times) obtained using the CIGALE outputs as described in the previous sections.
    \item For each stellar mass estimate, we used the 2D interpolation functions to obtain the mean and standard deviation of halo mass. Using these parameters, we produced 10 samples of halo mass values.
    \item Combining all the halo mass realizations for all redshift and stellar mass pairs (i.e. $1000\times100\times10=10^6$ total realizations), we finally produced estimates of DM for each galaxy halo intersecting the sightline (henceforth, \dmphotgal). \dmphotgal~values are calculated for each tuple of \mhalo~and \zphot~realizations as follows:
    \begin{enumerate}
        \item First the perpendicular distance from the FRB sightline is computed.
        \item Then, assuming the model for electron distribution as described in \citet{Simha+20}, \dmphotgal\ is estimated. We assumed that each halo extends to 1 virial radius  and the fraction of halo baryons present as hot ($>10^6$~K) gas is 0.75. This assumes that 25\% of the baryons in the galaxy is in condensed forms \citep[e.g. stars and neutral gas; see][]{Fukugita+98}. While this fraction may vary with halo properties \citep[e.g.][]{Behroozi+10},
        we emphasize that this is a relatively conservative maximal
        model for the CGM of galaxies, i.e.\ one
        may consider the estimates as upper limits.
    \end{enumerate}
\end{enumerate}
Finally, using the \dmphotgal~distributions for all galaxies in the sightline, we produced the distribution of their sum, i.e. \dmhalosphot.

In the fifth step we imposed some bounds on \mhalo~estimates to ensure reasonable values. Namely, the \mhalo\ estimated at a particular redshift grid point may not exceed $10^{12.8}~\mmsun$, which is nearing a typical galaxy group halo mass. Exceeding this value is allowed by the uncertainty limits from the SHMR. Therefore, we artificially capped the halo mass estimates to $10^{12.8}~\mmsun$, i.e. any halo mass realization above this limit was set by hand to $10^{12.8}~\mmsun$. Our \dmhalosphot\ distribution was largely unaffected by this choice of the upper limit as an overwhelming majority of galaxies (including the ones within  $2\arcmin$ to the sightline) have halo mass estimates much less than this limit. Additionally, it is often the case that the posterior distribution of \zphot~peaks beyond the FRB host redshift, \zzfrb = \zfrb. Even in this case there is a non-zero probability of the galactic redshift being below \zzfrb. For all \zphot~realizations beyond \zzfrb, we set \dmphotgal=$0~\mdmunits$. 

\begin{figure*}[!tb]
    \centering
    \includegraphics[width=\textwidth]{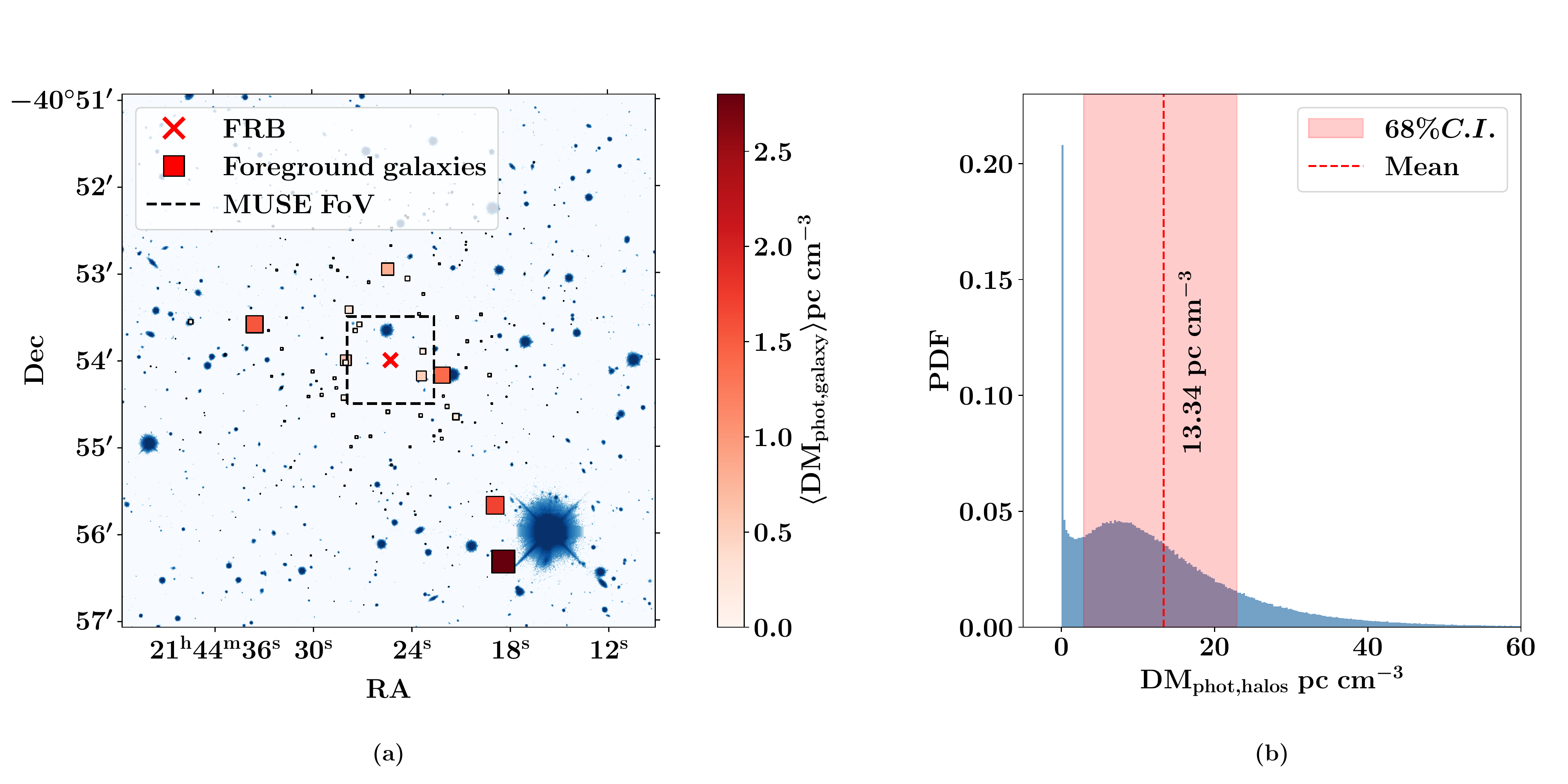}
    \caption{\footnotesize (a) Locations of DES galaxies (excluding those with MUSE and LRIS redshifts) coloured by their average estimates of \dmphotgal. Both the colors and the sizes of the points are proportional to the mean \dmphotgal.  The background image in blue is the DES $r$-band image of the field. The objects that fall within the MUSE field of view (black, dashed rectangle) do not have spectroscopic redshifts as their spectra did not have identifiable spectral features. (b) A realization of the PDF of \dmhalosphot\ estimated by producing $10^6$ realizations of the sum of \dmphotgal\ for all non-spectroscopic galaxies. The histogram counts are normalized to add up to unity. The large spike at 0 \dmunits\ is indicative of the possibility of most of these galaxies being in the background according to their \zphot\ posteriors.}
    \label{fig:avg_DM_sky}
\end{figure*}

\section{Results}
\label{sec:results}
Figure \ref{fig:avg_DM_sky}a shows the average \dmphotgal~contributed by each of the foreground sources estimated using this method. We excluded all sources for which we have spectroscopic redshifts from MUSE or LRIS. There were $\sim11000$ DES galaxies in our catalog. We had expected correctly that a large fraction of these sources do not contribute to \dmhalos. 

Based on these results, we targeted the 5 sources with highest mean \dmphotgal\ using Keck/LRIS ($\S$~\ref{sec:spectra}
and we detected line emission from one ($z = 0.07221$) of them, thus solidifying its redshift (listed in 
the last row of Table \ref{tab:muse_gals}). 

Figure \ref{fig:avg_DM_sky}b shows a realization of the final PDF of \dmhalosphot\ estimated from the 422 galaxies that have a non-zero probability of contributing to the FRB DM. Its mean value is $13$~\dmunits\ and the 68\% confidence bounds are $4$ and $23$~\dmunits. The spike at $0$~\dmunits\ arises from the fact that most galaxies have their redshift posterior distributions peaking beyond the FRB redshift, i.e.\ the majority of these are likely to have zero
contribution to \dmfrb.

Our sample of foreground galaxies with spectroscopic redshifts consists of three galaxies: two from our MUSE datacube and one from our LRIS pointings.  One of the MUSE galaxies (z=0.2859) does not have DES/WISE photometry. Therefore, we derived the stellar mass estimate from a pPXF \citep{ppxf} fit to its MUSE spectrum and assumed an error of 0.3 dex for $\log\mmstar/\mmsun$. 
From these galaxies, we estimated the mean net DM contribution of $7$~\dmunits with 68\% confidence bounds being 3 \dmunits~and 12 \dmunits. The bounds were estimated by propagating the uncertainties in the stellar mass and SHMR as described previously but the redshift is fixed.

\begin{figure}[!t]
    \centering
    \includegraphics[width=0.45\textwidth]{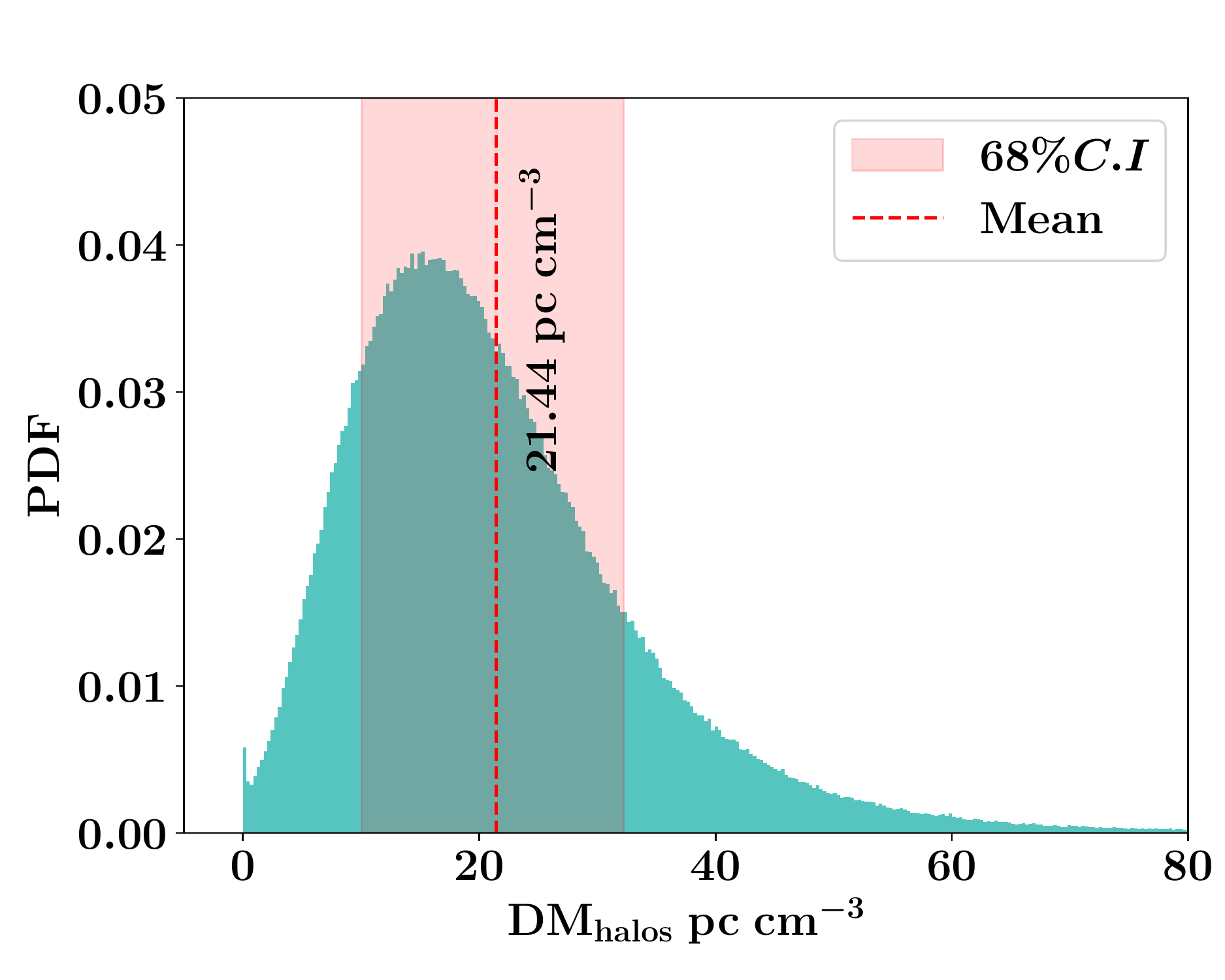}
    \caption{\footnotesize The distribution of \dmhalos~estimated using the full sample of foreground galaxies i.e. including galaxies with spectroscopic redshifts from MUSE and LRIS. The galaxies with spectra add 7.1 \dmunits~to \dmhalos~on average, thus shifting the mean value from 13.34 \dmunits~in Figure \ref{fig:avg_DM_sky}b to 21.44 \dmunits.
    }
    \label{fig:dm_halos_phot_spec}
\end{figure}

Thus, the mean \dmhalos\ estimate, which is the sum of the estimates from the two disjoint samples is 21 \dmunits\ and the 68\% confidence limits are 9 \dmunits~and 32 \dmunits. 
The full distribution is shown in 
Figure~\ref{fig:dm_halos_phot_spec}. 

\begin{figure}[!t]
    \centering
    \includegraphics[width=0.45\textwidth]{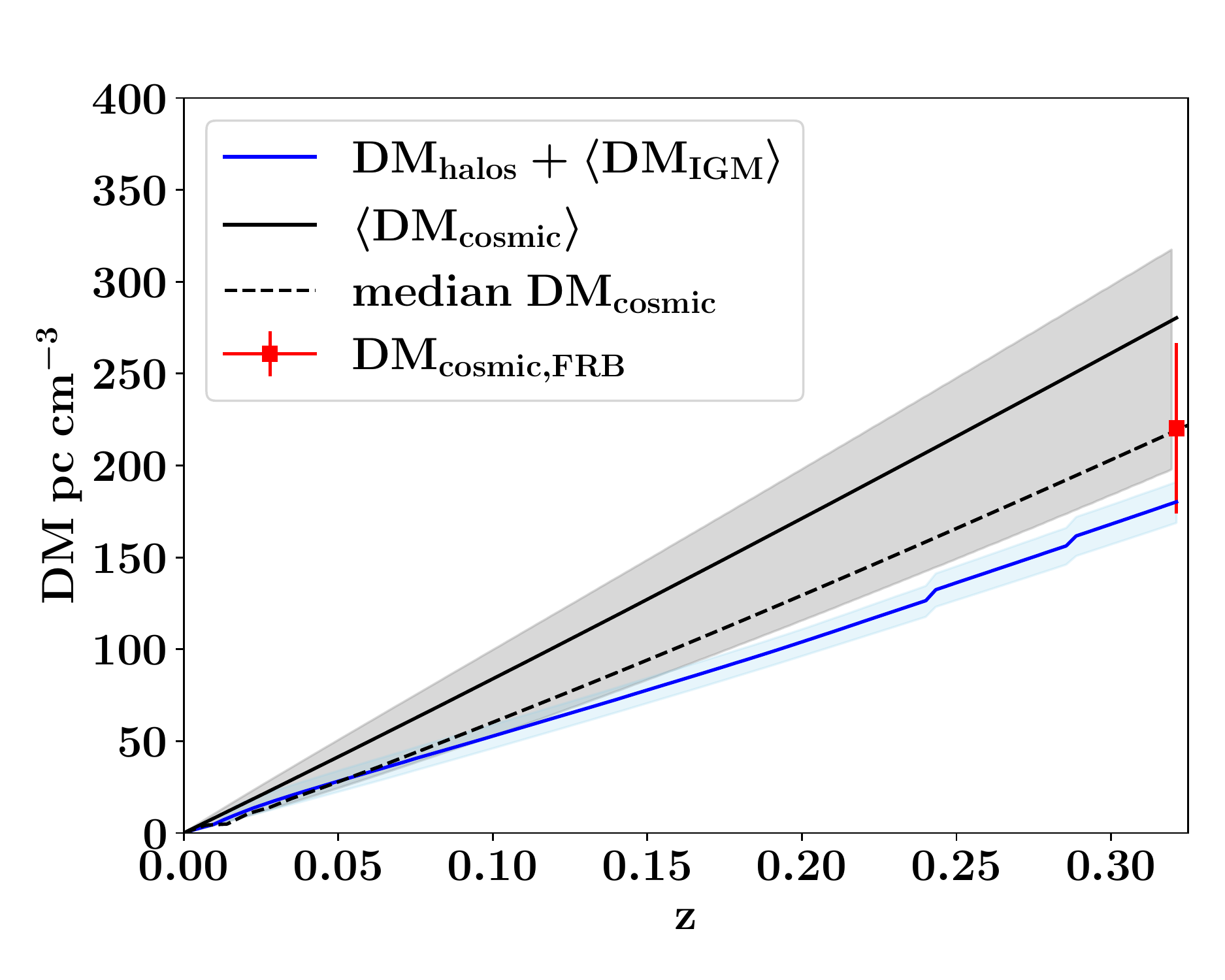}
    \caption{\footnotesize An estimate of \dmcosmic~for the FRB 180924 sightline (blue, solid line) which is a sum of \dmhalos~from this analysis and the average diffuse IGM contribution, \dmaigm. Starting from z = 0, \dmhalos~increases as one encounters halos along the sightline and the value at \zzfrb~is the one estimated in Figure~\ref{fig:dm_halos_phot_spec}. \dmaigm~increases similarly as more matter is met on average going out towards the FRB. The blue, shaded region corresponds to the 68\% confidence interval obtained from the uncertainty in \dmhalos. The red point at 220 \dmunits~is an estimate of $\rm DM_{\rm cosmic, FRB}$~ obtained by subtracting the Milky Way and host galaxy contributions from the net DM. The error bar is the net uncertainty in this estimate and corresponds to 50 \% uncertainty in each of the subtracted quantities, added in quadrature. The black, solid line is \dmacosmic~described by the Macquart relation and the gray shaded region represent its scatter ($1\sigma$ limits) due to the filamentary nature of the cosmic web. The black dotted line is the locus of all the median values of \dmcosmic~obtained from the same distribution \citep{macquart2020}.}
    \label{fig:dm_cosmic}
\end{figure}

\section{Discussion and concluding remarks}
\label{sec:discuss}

For FRB~190608, \citet{Simha+20} estimated \dmhalos~to be between 7 \dmunits~and 28 \dmunits. This corresponded to between 2\% and 8\% of the net DM and between 5\% and 20\% of \dmcosmic. In the case of FRB 190608, the theoretical average value, \dmahalos\ at \zzfrb\ is 44 \dmunits, a few times larger than the estimated \dmhalos in that sightline. This expectation value is computed assuming $\Lambda$CDM cosmological parameters from \citet{Planck15}, a model for the gas density in halos (the same as we have used previously), and the Aemulus halo mass function \citep[HMF; ][]{HMF}. The HMF is integrated between $10^{10.3}~\mmsun$ and $10^{16}~\mmsun$.

In the case of FRB 180924, the expected \dmahalos~is 121~\dmunits\ because it is more distant than than FRB 190608. Compared to this, the mean value of \dmhalos~ estimated in the previous section is 
just 21 \dmunits\ assuming the same CGM model. Thus, \dmhalos~is conclusively lower than average for this sightline, much like FRB 190608.

Figure \ref{fig:dm_cosmic} shows, with a solid blue line, the sum of our \dmhalos~estimate and \dmaigm,the average DM contribution of the diffuse IGM. We define \dmaigm~as \dmahalos, as computed above, subtracted from \dmacosmic, i.e. the mean Macquart relation ($\mdmaigm \equiv \mdmacosmic -\mdmahalos$). Comparing it to the $\rm DM_{\rm cosmic, FRB}$~estimate (shown as a red point with errors) we see that the two independently computed estimates are indeed consistent. Favored models of the cosmic web indicate that most FRB sightlines in the universe will have few if not zero dark matter halos intersecting them proximally \citep[e.g. ][]{macquart2020}.i.e. if one were to connect the median values of the \dmcosmic~distributions at each redshift, the resulting curve, which can be called the median Macquart relation, lies below the mean curve and is shown as the dotted, black line. Indeed, $\rm DM_{\rm cosmic, FRB}$ is coincident with this median curve. We therefore conclude that FRB 180924 is one such sightline.

We note here that there are indeed other models of gas distribution in the CGM, some of which predict larger dispersion measures, by a factor of a few, from individual halos, \citep[e.g. see Figure 1 of ][]{xyz19}. If we were to use any of these models which predict systematically higher DM contributions, both \dmhalos~and \dmahalos~would increase by the same factor, and therefore \dmhalos\ for this sightline would still be lower than average. Simultaneously, our estimate for \dmcosmic~in Figure 6 (the blue line) would decrease when using these models. This is because \dmaigm~constitutes the majority of the \dmcosmic~estimate and by definition, it decreases with increasing \dmahalos. One must be cautious when performing this exercise however. For instance, with our chosen model of halo gas distribution, we estimate \dmahalos~= 121 \dmunits. Since \dmacosmic~= 280 \dmunits~is independent of this model, doubling \dmahalos~would only leave $\sim40~\mdmunits$ for \dmaigm. This is low and likely unrealistic at the host redshift, especially compared to the \dmigm~estimate using the MCPM method for the FRB~190608 sightline by \citet{Simha+20}. Thus, to truly estimate \dmcosmic\ one cannot simply use \dmaigm , and a detailed, semi-empirical model of the cosmic web density is required.

In summary, we have shown that photometric data can be used effectively to constrain \dmhalos. While the uncertainty in this endeavor is significant, one can use this as a first step in identifying targets for efficient spectroscopic follow up observations. Having full spectroscopic coverage of the field is undeniably better as the photo-$z$ analysis can misidentify background sources as being in the foreground and vice-versa. In the near future, we intend to obtain spectra of field galaxies within a few degrees of FRB 180924 and perform a full cosmic web analysis, including a direct accounting of the diffuse IGM DM contribution. With upcoming large-scale spectroscopic surveys such as DESI, more FRB fields will have galaxies with precise redshifts and a statistical analysis of multiple FRB fields to constrain cosmic web properties such as the fraction of cosmic baryons will be enabled.

\textit{Acknowledgments:}

Authors S.S., N.T., J.X.P., and K.G.L. as members of the Fast and Fortunate for FRB
Follow-up team, acknowledge support from 
NSF grants AST-1911140 and AST-1910471.
K.G.L. acknowledges support from JSPS KAKENHI Grants
JP18H05868 and JP19K14755.
K.W.B. acknowledges Australian Research Council (ARC) grant DP180100857.
 N.T. acknowledges support by FONDECYT grant 11191217.
 S.C. gratefully acknowledges support from Swiss National Science Foundation grant PP00P2\_190092 and from the European Research Council (ERC) under the European Union’s Horizon 2020 research and innovation programme grant agreement No 864361.
 R.M.S. acknowledges support through Australian Research Council Future Fellowship FT190100155 and Discovery Project DP180100857. 
The Australian Square Kilometre Array Pathfinder is part of the Australia Telescope National Facility which is managed by CSIRO. 
Operation of ASKAP is funded by the Australian Government with support from the National Collaborative Research Infrastructure Strategy. ASKAP uses the resources of the Pawsey Supercomputing Centre. Establishment of ASKAP, the Murchison Radio-astronomy Observatory and the Pawsey Supercomputing Centre are initiatives of the Australian Government, with support from the Government of Western Australia and the Science and Industry Endowment Fund. 
We acknowledge the Wajarri Yamatji as the traditional owners of the Murchison Radio-astronomy Observatory site. 

Based on observations collected at the European Southern Observatory under ESO programmes 2102.A-5005 and 0104.A-0411. Spectra were also obtained at the W. M. Keck Observatory, which is operated as a scientific partnership among Caltech, the University of California, and the National Aeronautics and Space Administration (NASA). The Keck Observatory was made possible by the generous financial support of the W. M. Keck Foundation. The authors recognize and acknowledge the very significant cultural role and reverence that the summit of Mauna Kea has always had within the indigenous Hawaiian community. We are most fortunate to have the opportunity to conduct observations from this mountain.
\software{
EsoReflex \citep{esoreflex},
CubExtractor \citep{cubex},
SEP \citep{Barbary2016,SExtractor},
MARZ \citep{marz},
HMFEmulator \citep{HMF},
EAZY, \citep{eazy},
CIGALE \citep{cigale},
Astropy \citep{astropy:2018},
Numpy \citep{numpy},
Scipy \citep{scipy},
Matplotlib \citep{Hunter:2007}.
}

The Python scripts used to perform our analysis are available in our FRB GitHub repository (https://github.com/FRBs/FRB). 

\bibliography{frb180924_fg.bib}

\end{document}